\documentclass{PoS}

\title{Radio-Loud QSOs and Sub-Millimeter Galaxies: Space Distributions}

\ShortTitle{RQSOs and SMGs: space distributions}

\author{\speaker{J V Wall}\\
        Department of Physics and Astronomy, University of British Columbia, Vancouver,
B.C. V6T\,1Z1, Canada\\
        E-mail: \email{jvw@astro.ubc.ca}}

\abstract{A picture has emerged connecting QSOs with Sub-Millimetre 
Galaxies (SMGs)[1,2] through an evolutionary sequence in which forming galaxies are initially 
FIR-luminous but X-ray weak, similar to known SMGs. As the black hole and spheroid grow with 
time, the central QSO becomes powerful enough to terminate star formation and eject much 
of the fuel supply. The unobscured QSO activity subsequently declines to leave a quiescent 
spheroidal galaxy.  Here I describe parallel investigations of space density, one for 
a sample of radio-loud QSOs (RQSOs), and a second for SMGs. Each class shows both 
cosmic down-sizing and a redshift cutoff. The coincidence in apparent epoch 
of creation is marked; if it does not prove a causal connection, it is at least circumstantial 
evidence that the foregoing sequence is correct.

{\bf The  RQSO sample} was selected from the PKS 2.7-GHz survey over the southern hemisphere[3]; 
most of the 379 RQSOs of the complete sample have spectroscopic redshifts, and all are from 
2.7-GHz survey areas complete to flux densities ranging from 0.10 to 0.6~Jy. 
{\bf The SMG sample} comprises a complete set of 
35 objects from the GOODS-N supermap[4]; all have cross-waveband identifications. Most 
of the 35 have spectroscopic redshifts, the remainder, spectrophotometric redshift estimates. This 
is the first complete SMG sample for which space density analysis is possible.
}

\FullConference{From Planets to Dark Energy: The Modern Radio Universe\\
                 1$^{st}$-5$^{th}$ October 2007\\
                 The University of Manchester, United Kingdom}

\begin{document}

\begin{figure}[h]
  \begin{minipage}{7.5cm}
    \centerline{
      \includegraphics[angle=0,scale=0.35]{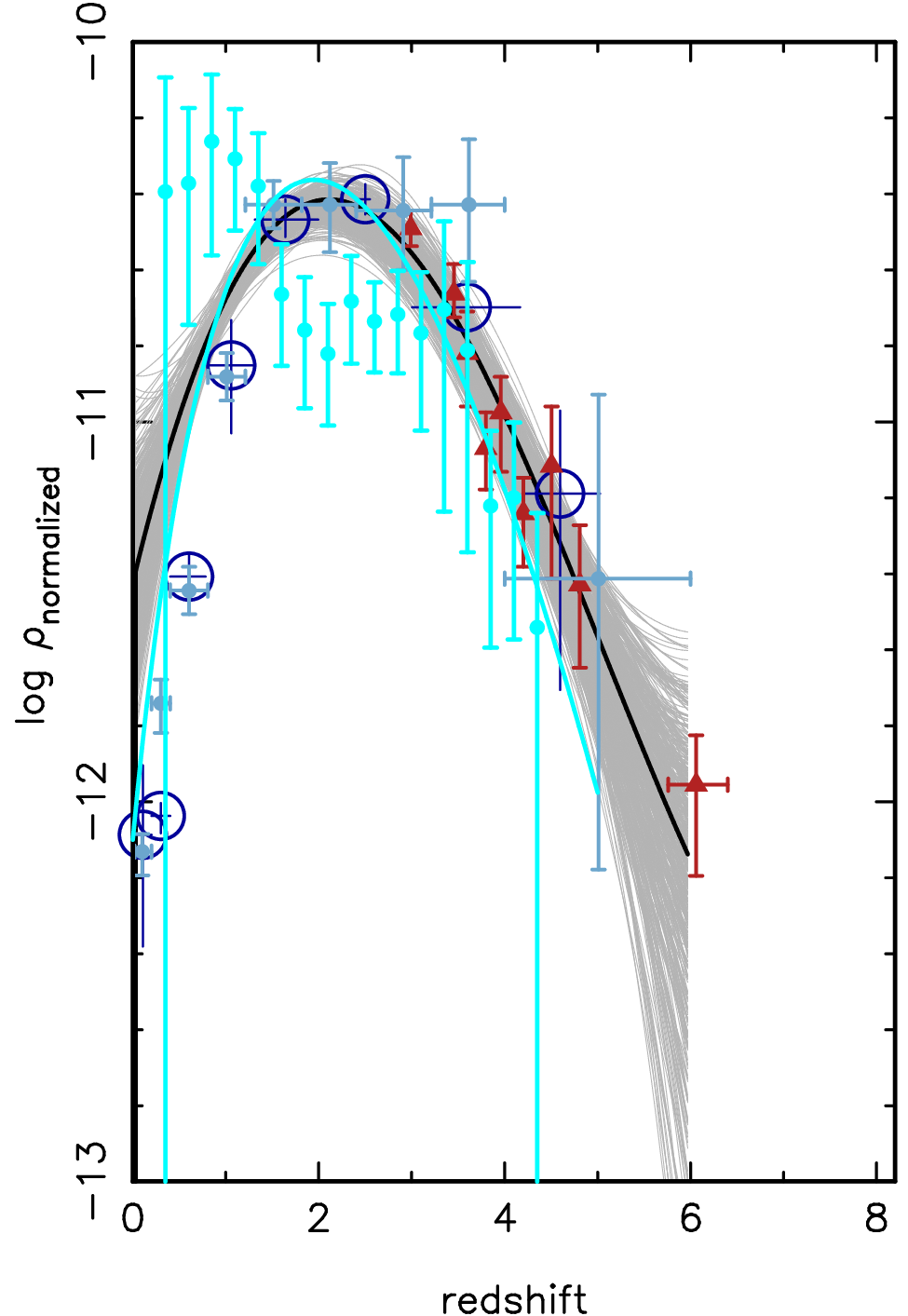}}
    \caption{Relative space density of RQSOs (black line, grey shading), X-ray QSOs 
(blue crosses, circles), and SMGs (turquoise points); see [5] for details.}
  \end{minipage}
\hfill
  \begin{minipage}{6.0cm}
    \centerline{
      \includegraphics[angle=-90,scale=0.27]{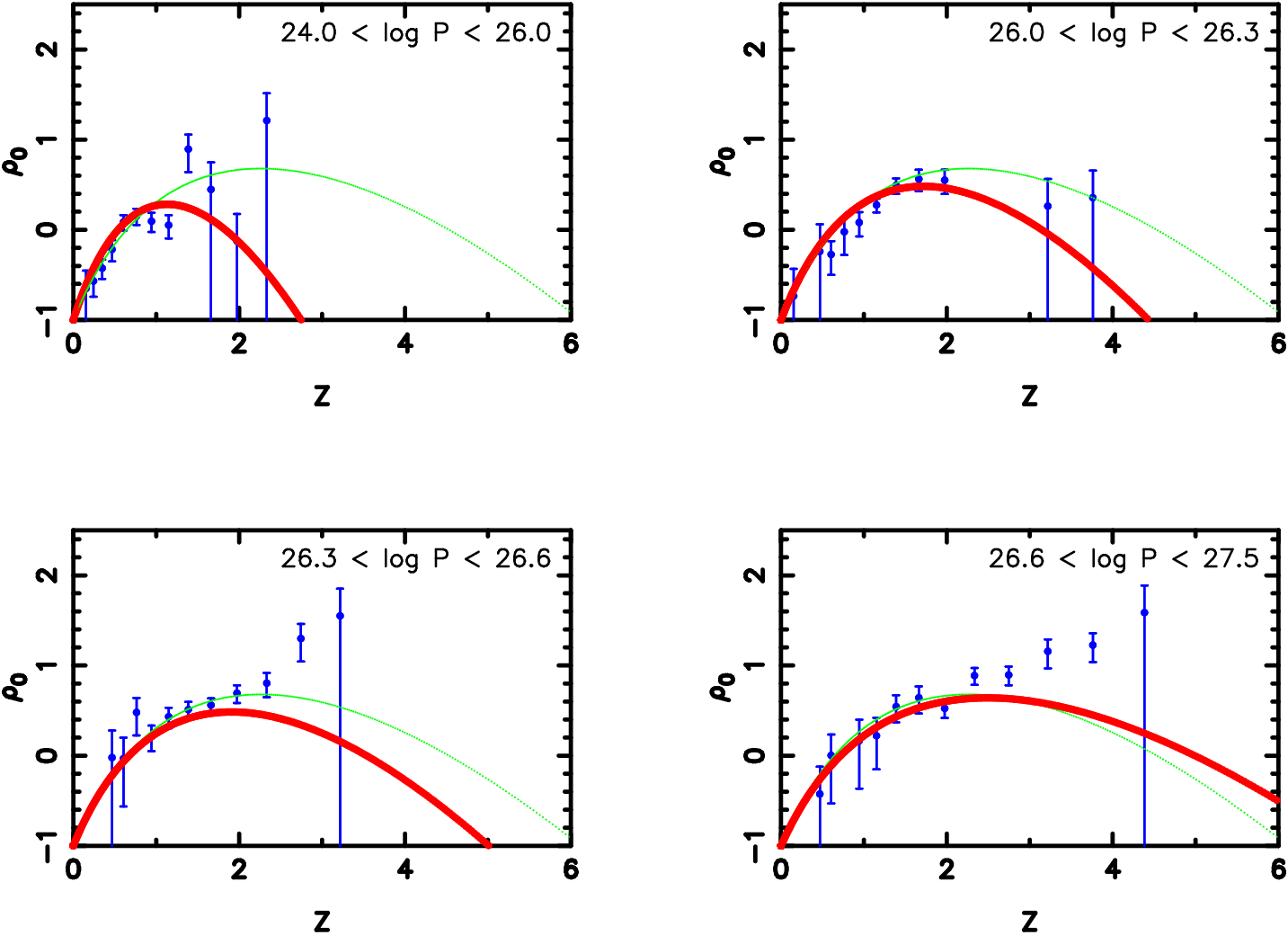}}
  \caption{Space density vs redshift for different RQSO luminosities, showing cosmic 
down-sizing for the RQSO 379 sample. A maximum likelihood analysis in 
luminosity bins (red curves) shows how the peak epoch of dominant activity moves to 
progressively lower redshifts as luminosity decreases [JVW, in preparation]. 
The faint green curve, repeated in 
each panel, shows the space density behaviour for the entire (unbinned) sample.
}
  \end{minipage}
\end{figure}

{\bf RQSO space density analysis} was carried out using a `single-source survey' 
technique[5], in which each sample member was considered as the result of its very own survey. 
In this manner, each source could be ascribed its particular survey flux limit and its own 
radio spectrum or K-correction, the two providing a unique cutoff line in the luminosity - 
redshift ($L - z $) plane. The contribution of each source to volume density could then be 
calculated.

Fig. 1 shows how  dramatic 
the density change is with redshift -- a rapid rise to the `quasar epoch' at redshift 1 to 2, 
followed by a decline in co-moving space density at redshifts beyond 3. This dramatic 
decline in space density at higher redshifts was subjected to bootstrap testing of
polynomial fits to combined sections of the luminosity functions; the significance level 
for this decline is below 0.001. Flux density variation had to be carefully considered [5].
To examine cosmic down-sizing, a maximum-likelihood modelling approach was needed, as used in
the  SMG analysis below. This down-sizing is conclusive, as shown in Fig. 2 [JVW, in preparation].

{\bf Analysis of the SMG sample} required a maximum-likelihood approach from the start; standard
$1/V_{\rm max}$ analyses will not work, as shown by the luminosity --
redshift plane of Fig.~3. Most of the SMGs would be visible at any $z$, i.e. $1/V_{\rm max} \approx
0$. We [6] combined the single-source survey technique with a maximum-likelihood 
analysis, assuming  (a)  Poisson distribution of the SMGs in cells of the $L-z$ plane, 
(b) a power law form for the luminosity function, and (c) an evolution function for changing 
the luminosity function with redshift of the form $(1+z)^{(k+\gamma z)}$. This form combines 
common power-law density evolution with a generic form of redshift cutoff if $\gamma$ should 
prove to be negative. We found the minimum in the likelihood function using 
a method of steepest descent. Fig.~4 encapsulates the results.
\begin{figure}[h]
  \begin{minipage}{7.0cm}
    \centerline{
      \includegraphics[angle=-90,scale=0.30]{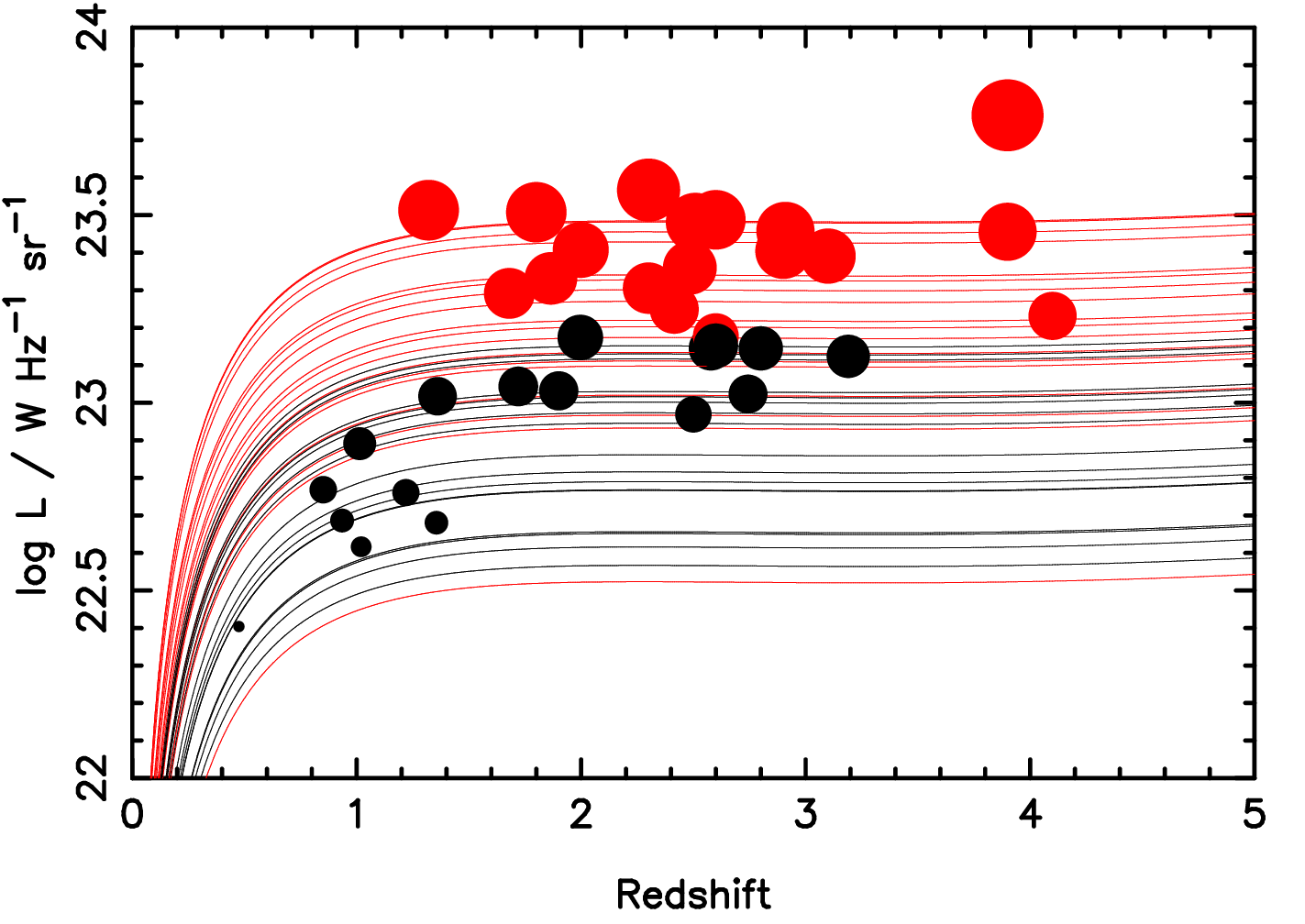}}
    \caption{Luminosity-redshift plane for the GOODS-N 35 SMG sample. The diameter of each point 
is proportional to log(L), and each line indicates a survey cutoff limit for one of the SMGs. Red
points: log $(L_{850\mu m}/{\rm W Hz}^{-1}{\rm sr}^{-1}) > 23.17$; 
black: log $(L_{850\mu m}/{\rm W Hz}^{-1}{\rm sr}^{-1}) < 23.17$.} 
  \end{minipage}
\hfill
  \begin{minipage}{7.0cm}
    \centerline{
      \includegraphics[angle=-90,scale=0.28]{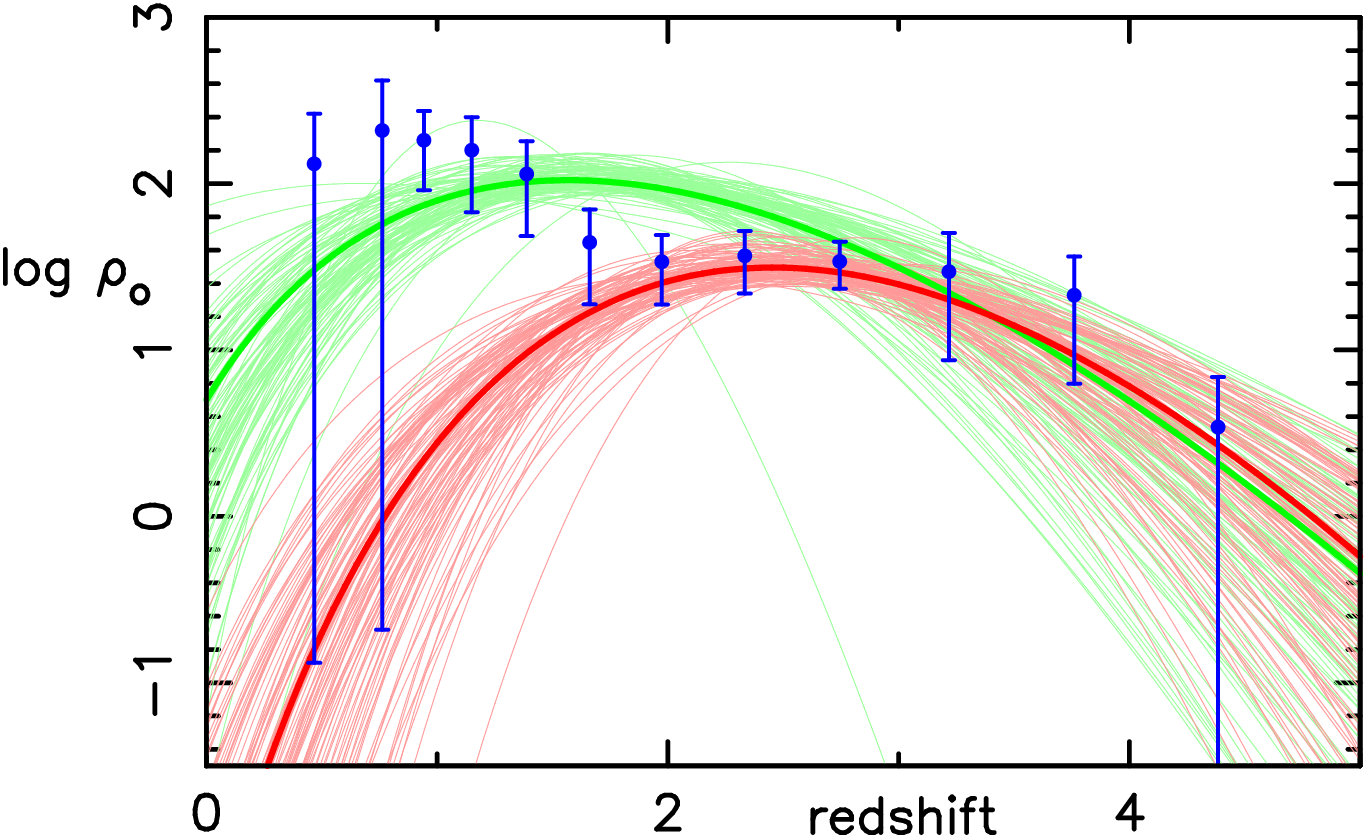}}
  \caption{SMG evolution. The points represent space densities over narrow redshift ranges,
assuming constant space density within these ranges. The best-fit 
parameters yield the solid green curve for the low-luminosity half of the sample and the red curve 
for the high-luminosity half. This splitting of the sample produced a very significant 
improvement in the overall fit to space densities -- see text.}
  \end{minipage}
\end{figure}

The redshift cutoff parameter $\gamma$ is negative at a significance level too small to 
measure. There is a redshift 
cutoff for the SMGs, as their distribution in Fig.~3 suggests. The overall behaviour of 
SMG space density is shown by the turquoise points in Fig.~1 and the blue points in Fig.~4.

We found (Fig.~4) a far more successful fit in modelling with {\it two SMG sub-populations}, 
divided (a priori) at the median sample luminosity. This result came as a surprise from 
bootstrap testing, which persistently showed bimodal distributions in probability planes of the
model parameters. The higher-luminosity component shows  much stronger evolution and a steeper 
slope for the luminosity function. This indicates cosmic down-sizing for SMGs.

\vspace{2.5mm}

\noindent{\bf Both populations demonstrate:}

1. Similar cosmological evolution, with a steep rise in space density to $z=1.5$, followed by 
a plateau to $z \sim 3$; and

2. A diminution in space density (level of significance <0.001) beyond z = 3. The form of 
this density diminution is virtually coincident as shown in Fig. 1. At minimum this 
is circumstantial evidence 
that the objects are related, may have a common origin, and may be linked in an 
evolutionary sequence such as that described in the introduction.

The further and surprising feature of the SMG space analysis is that the objects appear to 
constitute {\it two populations divided by luminosity}. Perhaps these sub-populations are 
the high-luminosity equivalents of nearby LIRGS and ULIRGS which are 
known to differ dramatically in evolution properties, space densities and epoch-dependent 
star-formation rates[7]. Large SCUBAII SMG samples will clarify these preliminary results.

\vspace{5mm}

\noindent{\bf References: }[1] Sanders D.B. et al. 1988, ApJ Lett 328, L35; 
[2] Alexander D.M. et al. 2005, Nat 434, 738; 
[3] Jackson C. et al. 2002, A\&A 386, 97; [4] Pope A. et al. 2005, MNRAS 358, 149; 
ibid 2006, MNRAS 370, 1185; [5] Wall J.V. et al. 2005, A\&A 434, 133; [6] Wall J.V. et al. 
2007, astro-ph/0702682, accepted MNRAS; [7] Chary R-R. 2006, astro-ph/0612736



\end{document}